\documentclass[12pt,preprint]{aastex}

\shorttitle{DNC/HNC in G34.43+00.24 MM3}
\shortauthors{Sakai et al.}

\begin{document}


\title{ALMA Observations of the IRDC Clump G34.43+00.24 MM3: DNC/HNC Ratio}


\author{Takeshi Sakai\altaffilmark{1}, Nami Sakai\altaffilmark{2}, Kenji Furuya\altaffilmark{3}, Yuri Aikawa\altaffilmark{4}, Tomoya Hirota\altaffilmark{5,6}, Jonathan B. Foster\altaffilmark{7}, Patricio Sanhueza\altaffilmark{5}, James M. Jackson\altaffilmark{8}, and Satoshi Yamamoto\altaffilmark{2}}

\altaffiltext{1}{Graduate School of Informatics and Engineering, The University of Electro-Communications, Chofu, Tokyo 182-8585, Japan.}
\altaffiltext{2}{Department of Physics, Graduate School of Science, The University of Tokyo, Tokyo 113-0033, Japan.}
\altaffiltext{3}{Leiden Observatory, Leiden University, P.O. Box 9513, 2300 RA Leiden, The Netherlands}
\altaffiltext{4}{Department of Earth and Planetary Sciences, Kobe University, Kobe 657-8501, Japan.}
\altaffiltext{5}{National Astronomical Observatory of Japan, Osawa, Mitaka, Tokyo 181-8588, Japan.}
\altaffiltext{6}{Department of Astronomical Sciences, Graduate University for Advanced Studies, Mitaka, Tokyo 181-8588, Japan.}
\altaffiltext{7}{Yale Center for Astronomy and Astrophysics, Yale University, New Haven, CT 06520, USA.}
\altaffiltext{8}{Institute for Astrophysical Research, Boston University, Boston, MA 02215, USA.}


\begin{abstract}
We have observed the clump G34.43+00.24 MM3 associated with an infrared dark cloud in DNC $J$=3--2, HN$^{13}$C $J$=3--2, and N$_2$H$^+$ $J$=3--2 with the Atacama Large Millimeter/submillimeter Array (ALMA).  
The N$_2$H$^+$ emission is found to be relatively weak near the hot core and the outflows, and its distribution is clearly anti-correlated with the CS emission. This result indicates that a young outflow is interacting with cold ambient gas.
The HN$^{13}$C emission is compact and mostly emanates from the hot core, whereas the DNC emission is extended around the hot core. Thus, the DNC and HN$^{13}$C emission traces warm regions near the protostar differently.
The DNC emission is stronger than the HN$^{13}$C emission toward most parts of this clump.
The DNC/HNC abundance ratio averaged within a 15$^{\prime\prime}$$\times$15$^{\prime\prime}$ area around the phase center is higher than 0.06. This ratio is much higher than the value obtained by the previous single-dish observations of DNC and HN$^{13}$C $J$=1--0 ($\sim$0.003).
It seems likely that the DNC and HNC emission observed with the single-dish telescope traces lower density envelopes, while that observed with ALMA traces higher density and highly deuterated regions.
We have compared the observational results with chemical-model results in order to investigate the behavior of DNC and HNC in the dense cores.
Taking these results into account, we suggest that the low DNC/HNC ratio in the high-mass sources obtained by the single-dish observations are at least partly due to the low filling factor of the high density regions.

\end{abstract}





\section{Introduction}

It is estimated that 70-90 \% of stars in the galaxy are born in clusters (Lada \& Lada, 2003). Since molecular clumps (size $>$ 0.1 pc, mass $>$ several 10 $M_\odot$) are the birth places of clusters, understanding star formation in clumps is an important issue for astronomy.
Toward this goal, it is essential to investigate the initial physical conditions before the onset of cluster formation. 
Deuterium fractionation ratios could be useful for probing such past physical conditions of star-forming clumps in their starless phase.

In cold molecular clouds, the observed D/H abundance ratios in molecules (0.001-0.1; e.g., Watson 1974; van Dishoeck et al. 1995; Bergin \& Tafalla, 2007; Ceccarelli et al. 2014) is known to be much grater than the cosmic D/H ratio ($\sim$10$^{-5}$).
This is due to the exothermic isotope-exchange reactions, such as  
\begin{equation}
{\rm H}_3^+ + {\rm HD} \rightarrow {\rm H}_2{\rm D}^+ + {\rm H}_2\;,
\end{equation}
\begin{equation}
{\rm CH}_3^+ + {\rm HD} \rightarrow {\rm CH}_2{\rm D}^+ + {\rm H}_2\;,
\end{equation}
\begin{equation}
{\rm C}_2{\rm H}_2^+ + {\rm HD} \rightarrow {\rm C}_2{\rm HD}^+ + {\rm H}_2\;.
\end{equation}
Since their backward reactions are endothermic, the equilibrium deuterium fractionation ratios of molecules depend on temperature, where lower temperatures lead to higher deuterium fractionation ratios. 
In addition, depletion of CO onto dust grains accelerates the deuterium fractionation below $\sim$20 K, because it extends the lifetime of H$_3^+$ and H$_2$D$^+$. 
In fact, high deuterium fractionation has been reported in evolved low-mass prestellar cores (e.g., Caselli et al. 1999; Hirota et al. 2011).
Ortho-to-para ratio of H$_2$ also affects the molecular D/H ratio. The lowest state of ortho-H$_2$ lies $\sim$170 K above the ground state of para-H$_2$, which helps to overcome the endothermicity of the backward reaction of (1)　(Flower et al. 2006; Pagani et al. 2011).

After the birth of a protostar, deuterium fractionation ratios start to decrease toward the equilibrium values at the elevated temperature.
The timescale for this change depends on the molecular species, and is different between ionic and neutral species.
For example, the main destruction pathways for DNC are reactions with molecular ions, such as DNC $+$ HCO$^+$ $\to$ DNCH$^+$ $+$ CO and DNC $+$ H$_3^+$ $\to$ DNCH$^+$ $+$ H$_2$.
Then the dissociative recombination produces CN, although some fraction of DNCH$^+$ goes back to DNC.
The timescale for these reactions is typically 10$^{3-4}$ yr. 
Thus the initial DNC/HNC abundance ratio is almost conserved for this timescale after the onset of star formation.
This is in contrast to the ionic species, which are destroyed by electron recombination in a relatively short timescales ($<$100 yr).
For N$_2$D$^+$ (N$_2$H$^+$ as well), the reaction with CO is the main destruction process above the sublimation temperature of CO ($\sim$20 K), whose timescale is 1-10 yr.

The behavior of the deuterium fractionation in high temperature regions was discussed by many authors (Plambeck \& Wright 1987; Rodgers \& Miller 1996; Charnley et al. 1997; Albertsson et al. 2013; Taquet et al. 2014).
Aikawa et al. (2012), for example, investigated in detail the evolutionary change in the deuterium fractionation ratios from low-mass prestellar cores to protostellar cores, confirming the behavior of the neutral and ionic species.
Thus, for the neutral species, the deuterium fractionation ratios of active star forming regions reflect those of their starless phase just before the onset of star formation.
Therefore, we can extract information of a cold ``starless'' phase from observations of ``star-forming'' regions.

With the above motivation in mind, Sakai et al. (2012) observed the DNC/HNC ratio toward 18 clumps, including infrared dark clouds (IRDCs) and high-mass protostellar objects (HMPOs), by using the Nobeyama Radio Observatory (NRO) 45 m telescope (beam size$\sim$20$^{\prime\prime}$).
They found that the DNC/HNC ratio of the high-mass sources is lower than that of the low-mass sources, and that the DNC/HNC ratio of some IRDCs is lower than that of the HMPOs.
These trends cannot be solely explained by the difference in the current temperature. They suggested that the DNC/HNC abundance ratio could reflect the initial condition of each source.
However, the spatial resolution of the single-dish observations was insufficient to resolve individual star-forming cores in the high-mass clumps.
In order to investigate the origin of the the low deuterium fractionation ratios of the IRDC clumps in more detail, high angular resolution observations are crucial.  In this paper, we observe G34.43+00.24 MM3, which has the lowest DNC/HNC ratio in the sample of our previous single-dish observations, at sub-arcsecond resolution with the Atacama Large Millimeter/submillimeter Array (ALMA).

G34.43+00.24 is a filamentary IRDC (Garay et al. 2004; Rathborne et al. 2005; 2006; Sanhueza et al. 2010), and contains 9 clumps (MM1-MM9: Rathborne et al. 2006). 
G34.43+00.24 MM3 is the third most massive clump in IRDC G34.43+00.24.
Recently, we observed this clump in the 1.3 mm continuum and several molecular lines with ALMA and in $K$-band (2.2 $\mu$m) with the Keck telescope (Sakai et al. 2013; Paper I).
These observations reveal a very young hot core/outflow system toward the center of this clump. We found that the SiO and CH$_3$OH emission is strong toward the hot core/outflow system (see Figure 1 in Paper I).
In addition, we also detected extended emission of SiO, CS, and CH$_3$OH, which is not associated with the hot core/outflow system (see Figure 4 in Paper I). 
We suggested that this emission may be related to past star formation activity in the clump.
Therefore, it is likely that this clump has already experienced active formation of low-mass stars in an early stage of clump evolution. 

The distance to this IRDC is estimated from VLBI parallax observations (1.56 kpc; Kurayama et al. 2011), and Sakai et al. (2013) used this distance. However, the distances estimated by assuming a Galactic rotation curve (kinematic distance: Sakai et al. 2008; Sanhueza et al. 2012) and by using the near infrared extinction method (Foster et al. 2012) (3-4 kpc) are different from the VLBI distance. This discrepancy is discussed in detail by Foster et al. (2012, 2014).
Even if a distance of 3.9 kpc is employed, the outflow in this clump has a dynamical age less than 1900 yr.
The conclusions of Sakai et al. (2013) have no change, if  the large distance is adopted.

In this paper, we present the observations of the DNC, HN$^{13}$C and N$_2$H$^+$ lines obtained with ALMA. We derive the DNC/HNC abundance ratio in the densest part of G34.43+00.24 MM3, and assess a relationship between star formation activity and the DNC/HNC abundance ratio. Furthermore, we reconsider the single-dish survey results from Sakai et al. (2012), taking the ALMA results into account.

\section{OBSERVATIONS}

We observed G34.43+00.24 MM3 with ALMA Band 6 (211-275 GHz) and Band 7 (275-373 GHz) on 2012 August 11, 15, and 26.
The phase center was (R.A.(J2000), Dec.(J2000)) = (18$^{\rm h}$53$^{\rm m}$20.4$^{\rm s}$, 1$^{\circ}$28$^{\prime}$23.0$^{\prime\prime}$).
The parameters for the Band 6 observations are described in Paper I. 
The Band 7 observations were carried out in the extended configuration with 23 antennas, providing baseline coverage from 20.1 m to 384.1 m.
The spectrometers were used with the 234 MHz mode, whose channel width is 61 kHz.
Since we averaged 8 channels for all the data in order to reduce the noise, the channel width shown in this paper is 488 kHz, corresponding to the velocity width of 0.64 km s$^{-1}$ at 230 GHz.
The bandpass calibration was carried out by observing J1924-292.
The flux calibration was carried out by observing Neptune and J1751+096.

The data were reduced by using the CASA software package.
The observed lines and their parameters, including the synthesized beam size for each line, are listed in Table 1. Although we observed the N$_2$D$^+$ $J$=3--2 line (231.32 GHz), the N$_2$D$^+$ emission is not detected anywhere in the field of view. This non-detection is partly due to the poor S/N ratio of the N$_2$D$^+$ data; since the N$_2$D$^+$ line overlaps a narrow atmospheric absorption line (O$_3$ 16$_{1,15}$--16$_{0,16}$; 231.281 GHz) the noise level was higher at this frequency.

\section{RESULTS}

\subsection{Integrated Intensity Maps}

Figures 1a-c show the integrated intensity maps of DNC $J$=3--2, N$_2$H$^+$ $J$=3--2, and HN$^{13}$C $J$=3--2.  In these figures, the distribution is clearly different from molecule to molecule.
The DNC emission (Figure 1a) peaks near the hot core (yellow star mark), and is extended toward the north of the hot core. A filamentary structure is also seen toward the south of the hot core.
In Figure 1b, the HN$^{13}$C emission is very compact, and it peaks toward the hot core.
No extended emission of HN$^{13}$C is detected.
On the other hand, the N$_2$H$^+$ $J$=3--2 emission is distributed throughout a relatively large area in this clump (Figure 1c).  
Near the hot core, the N$_2$H$^+$ emission shows a clumpy structure. 
A filamentary structure is seen in N$_2$H$^+$ toward the north of the hot core.  

In Figure 1d, we plot the CS $J$=5--4 (Paper I) and DNC contours superposed on the N$_2$H$^+$ color image.
In the eastern part of the map, we can see a collimated outflow traced by CS.
The CS emission seen in the southwestern part of the map should be outflows driven by embedded young stellar objects, as discussed in Paper I.  In this figure, the N$_2$H$^+$ emission is found to decrease toward the hot core and the outflow, while the DNC emission is detected near the hot core/outflow system.

Figure 2 shows the DNC and HN$^{13}$C contours superposed on the CH$_3$OH $J_K$=10$_3$--9$_2$ $A^-$ image (Paper I).
Since the upper state energy of CH$_3$OH $J_K$=10$_3$--9$_2$ $A^-$ is 165 K, this line traces the hot core.
In Figure 2, we can clearly see that the HN$^{13}$C emission come from the hot core and that the DNC peak is offset from the hot core.

\subsection{Channel Maps}

To investigate the relationship among the CS, N$_2$H$^+$, and DNC distributions, we present velocity-channel maps in Figure 3.
The CS emission appears over a wide velocity range, while the DNC and N$_2$H$^+$ emission appears only in a relatively narrow velocity range.
In the velocity range from 58 km s$^{-1}$ to 60 km s$^{-1}$, the N$_2$H$^+$ emission is seen on the northern edge of the CS outflow lobe. Toward this region, the N$_2$H$^+$ and CS distributions are clearly anti-correlated.
This observational feature likely indicates that the outflow is interacting with the cold dense gas.  
This interpretation is also supported by detection of the class I CH$_3$OH maser toward the interacting region (Yanagida et al. 2014).

In Figure 3, the DNC emission is seen near the interacting region at the velocity of 58-60 km s$^{-1}$.  However, the DNC peak is clearly offset from the CS and N$_2$H$^+$ peaks. Since the DNC peak is rather close to the hot core, DNC seems abundant in warm gas around the protostar.  Furthermore, the offset of the DNC peak from the CS peak indicates that the DNC is not abundant in the interacting region. 

In the velocities of 59.27-59.86 km s$^{-1}$, a filamentary structure in the DNC emission is seen toward the south of the hot core.
Although the N$_2$H$^+$ emission is detected in some parts of the filamentary structure of the DNC emission, most of the filament is only traced by the DNC emission. 

\subsection{Spectra}

As seen in Figures 1 and 2, the DNC emission is stronger than the HN$^{13}$C emission in several positions inside the clump.
Figure 4b shows the spectra of DNC and HN$^{13}$C averaged within the 15$^{\prime\prime}$$\times$15$^{\prime\prime}$ region around the phase center, where no primary beam corrections are applied. 
In this figure, the DNC emission is stronger than the HN$^{13}$C emission.
This is apparently inconsistent with the result of the single-dish observations of DNC $J$=1--0 and HN$^{13}$C $J$=1--0 by Sakai et al. (2012), as shown in Figure 4a.  This discrepancy will be discussed later.

In Figures 4c and 4d, we present the DNC and HN$^{13}$C spectra toward the DNC and HN$^{13}$C peaks, respectively.
Note that these spectra are not averaged.
The DNC peak is about 0$^{\prime\prime}$.8 offset from the HN$^{13}$C peak.
Toward the DNC peak, the DNC emission is stronger than the HN$^{13}$C emission. The DNC spectrum shows a double peak, while the HN$^{13}$C spectrum shows a single peak. Since the velocity of HN$^{13}$C emission is coincident with the lower velocity component of the DNC emission, the double peak of the DNC emission corresponds to two different velocity components and is not a self-absorbed profile.

Toward the HN$^{13}$C peak, the HN$^{13}$C emission is stronger and broader than the DNC emission. In addition, the peak velocity of the HN$^{13}$C line is offset from that of the DNC line.
Since there are no other detectable molecular lines in this frequency range, except for the (CH$_3$)$_2$O lines seen around 75 km s$^{-1}$ (Figure 4d), the broad emission is only due to the HN$^{13}$C emission.
As mentioned above, the HN$^{13}$C peak almost coincides with the hot core position.  In Figure 4d, we also plot the CH$_3$OH $J_K$=$10_2$--$9_3$ $A^-$ spectrum (Paper I). The velocity range of the HN$^{13}$C emission is covered by that of the CH$_3$OH $J_K$=$10_2$--$9_3$ $A^-$ emission. Thus, the broad HN$^{13}$C emission is likely tracing the hot core. 
Since the DNC emission is narrow toward the hot core position, it is likely that the DNC emission comes from a colder envelope rather than the hot core.

\section{ANALYSIS AND DISCUSSION}

\subsection{Temperature of the N$_2$H$^+$ and DNC Emitting Regions}

The emitting region of N$_2$H$^+$ is found to be quite different from that of DNC. This suggests that these molecules trace different physical conditions.  It is known that N$_2$H$^+$ is less abundant in warm gas without CO depletion, because it is destroyed by reactions with CO.
Thus, the weak N$_2$H$^+$ emission near the hot core and the outflow indicates that the temperature is higher than the sublimation temperature of CO ($\sim$20 K). In contrast, the N$_2$H$^+$ emitting regions should be rather cold.

As for DNC, the main destruction mechanism is reactions with HCO$^+$ or H$_3^+$, and consequently, the destruction timescale of DNC is much longer than that of N$_2$H$^+$, as mentioned in Section 1.
Therefore, the DNC molecule is expected to survive at temperatures above 20 K for $\sim$10$^{3-4}$ yr.
Since the N$_2$H$^+$ emission is weak toward the DNC emitting regions, the DNC emission seems to trace relatively warm regions.

In addition, DNC may be formed on grain surfaces during the cold starless phase. The sublimation temperature of DNC and HNC is estimated to be about 80 K, if their binding energy is assumed to be the same as HCN ($E$/$k$=4170 K) reported in Yamamoto et al. (1983).
Thus, the abundances of HNC and DNC could increase in hot regions by sublimation from icy dust mantles.
As shown in Figure 2, the DNC peak is offset from the hot core, and no heating source is probably associated with the DNC peak.
The gas temperature of the DNC peak is therefore thought to be lower than that of the hot core.
If the DNC abundance is enhanced due to sublimation, the HN$^{13}$C abundance should be also enhanced.
However, the HN$^{13}$C emission is faint toward the DNC peak.
Thus, it is most likely that the large DNC abundance toward the DNC peak does not originate from the sublimation of icy dust mantles.
The temperature of the DNC emitting regions, except for the hot core, would be in a range from 20 K to 80 K.

\subsection{DNC/HNC Ratio}

We derive the DNC/HNC abundance ratio from the spectra shown in Figures 4b-d.
We obtain the abundance ratio by taking the ratio between the DNC column density, $N$(DNC), and the HNC column density, $N$(HNC).
The column densities were derived by assuming optically thin emission under local thermodynamic equilibrium conditions. For simplicity, we assume the same excitation temperature for DNC and HN$^{13}$C.
We calculate the column densities in the range previously defined: 20-80 K at the DNC peak and 20-130 K at the HN$^{13}$C peak position.

In order to estimate $N$(HNC) from $N$(HN$^{13}$C), Sakai et al. (2012) used the $^{12}$C/$^{13}$C ratio of 62, which is estimated from the CO and H$_2$CO data by Wilson \& Rood (1994). However, the HN$^{12}$C/HN$^{13}$C ratio may be higher than the $^{12}$CO/$^{13}$CO ratio. According to the model calculations by Furuya et al. (2011), the HN$^{12}$C/HN$^{13}$C ratio could be higher by a factor of 2 than the $^{12}$CO/$^{13}$CO ratio in dense ($n$$>$10$^5$ cm$^{-3}$) regions.  Hence, we assume the HN$^{12}$C/HN$^{13}$C ratio to be 60-80.
The derived DNC/HNC abundance ratios are listed in Table 2.

The DNC/HNC abundance ratio derived from the averaged spectra is higher than 0.06.  This is much higher than that obtained from the single-dish observations (0.003$\pm$0.001: Sakai et al. 2012).
The DNC/HNC ratio toward the DNC peak (0.03-0.08) is also higher than that of the single-dish observation.
These values are comparable to those observed in low-mass star forming regions ($\sim$0.06; Sakai et al. 2012, Hirota et al. 2001, 2011).

Toward the HN$^{13}$C peak, the derived DNC/HNC abundance ratio is 0.001-0.004. 
This ratio is comparable to that of the single-dish observation.
However, the emission from the hot core is not the main component producing the emission detected with the single-dish telescope, because the hot core is compact and the spectral shape toward the HN$^{13}$C peak is different from that of the single-dish observation (see Figure 4).

\subsection{Origin of the DNC/HNC Ratio}

The DNC/HNC ratio is different between the single-dish and ALMA observations.  
One possibility for the weak HN$^{13}$C emission is that the HN$^{13}$C emission is so extended that its most part is resolved out in the ALMA observations.
However, the HN$^{13}$C emission should be detected toward the ``compact" DNC emitting regions even in this case, as long as the DNC/HNC ratio is relatively low ($\sim$0.003). 
Moreover, the critical density of the HN$^{13}$C $J$=3--2 emission is as high as $\sim$10$^7$ cm$^{-3}$, and it is unlikely that such high density regions are extended throughout the clump.
Therefore, we do not think that the resolved-out problem is the main reason for the weak HN$^{13}$C $J$=3--2 emission in this ALMA observation.

The most probable explanation for the different DNC/HNC ratios is that the HN$^{13}$C emission traces different density regions between the single-dish and ALMA observations.
In the single-dish observations, we observed the $J$=1--0 line.
The critical density of HN$^{13}$C and DNC $J$=1--0 ($\sim$10$^5$ cm$^{-3}$) is lower than that of HN$^{13}$C and DNC $J$=3--2 ($\sim$10$^7$ cm$^{-3}$). Therefore the single-dish observations preferentially trace the lower density envelopes, which may have a relatively low DNC/HNC ratio.
Fontani et al. (2014) also pointed out that the DNC/HNC ratio derived from the $J$=1--0 data traces the regions with a density of $\sim$10$^4$ cm$^{-3}$, based on the comparison between their single-dish observations and chemical model calculations.
On the other hand, the ALMA observations mainly trace denser regions with higher deuterium fractionation ratios.
The origin of the low DNC/HNC ratio in the diffuse regions will be discussed in the next section.

In Figure 5, we summarize our interpretation of the results in a schematic illustration.
The clump consists of extended inter-core medium ($\sim$10$^{4-5}$ cm$^{-3}$) with embedded dense cores ($\sim$10$^{6-7}$ cm$^{-3}$).
The DNC/HNC abundance ratio is relatively high in the densest regions, with the exception of the hot core.
Since the filling factor of the dense regions is relatively low for the single dish $J$=1--0 observation, the DNC/HNC ratio of the single dish observation is lower than that of the interferometer observations.　The filling factor of the DNC emitting region within the 15$^{\prime\prime}$$\times$15$^{\prime\prime}$ area is estimated to be about 0.1.
Toward the hot core, the DNC/HNC ratio is rather low as compared with the nearby dense regions. The low DNC/HNC ratio in the hot core will be discussed in the next section.

\subsection{Comparison with Chemical Model Calculations}

To investigate the behavior of the DNC/HNC abundance ratio in more detail, we carried out model calculations of gas-phase and grain-surface chemistry. We adopt the chemical reaction network of Furuya et al. (2015, in prep.); it is originally based on the reaction network of Garrod \& Herbst (2006) and high-temperature gas-phase reaction network of Harada et al. (2010, 2012), and has been extended to include multideuterated species (Aikawa et al. 2012; Furuya et al. 2013) and nuclear spin states of H$_2$, H$_3^+$, and their deuterated isotopologues (Hincelin et al. 2015, in prep.; Coutens et al. 2014). The rate coefficients of reactions between CH$_3^+$ + H$_2$ and their deuterated isotologues are updated referring to Roueff et al. (2013). We adopt a three-phase model, which consists of gas, chemically active ice mantle, and inert ice mantle (Hasegawa \& Herbst 1993); in the ice mantle, chemical reactions occur only in four layers from the surface  (Vasyunin \& Herbst 2013). 
Swapping between the surface active layer and innert ice mantle, i.e. the thermal diffusion of ice mantle, is included following Garrod (2013).
The cosmic ray ionization rate is set to be 2.6$\times$10$^{-17}$ s$^{-1}$ (van der Tak \& van Dishoeck 2000). Note that the binding energy of HNC used here (4170 K) is higher than that in Sakai et al. (2012; 2050 K); the former is based on the measurement of vapor pressure of HCN (Yamamoto et al. 1983), while the latter value is adopted from Garrod \& Herbst (2006). Thus, the sublimation temperature of HNC is $\sim$80 K in the current model, while that was $\sim$40 K in the model of Sakai et al. (2012).

First, we assume a uniform cloud with constant temperature and constant density of 10$^4$ cm$^{-3}$ or 10$^6$ cm$^{-3}$, and calculate the temporal variation of chemical compositions. Visual extinction was assumed to be 10 mag, so that photoreactions induced by the interstellar radiation field are unimportant. As an initial condition, all the elements are assumed to be in the form of neutral atom or atomic ion, depending on their ionization potentials, except for hydrogen and deuterium, which are in molecular form (H$_2$ and HD). The o/p ratio of H$_2$ affects the molecular D/H ratio, because the internal energy of ortho-H$_2$ acts as a reservoir of chemical energy which hampers the deuteration of H$_3^+$. Recently, Brunken et al. (2014) measured the o/p ratio of H$_2$D$^+$, and derived the o/p H$_2$ ratio to be $\sim$2$\times$10$^{-4}$ towards IRAS 16293-2422. It is not clear, however, what the optimal initial value of o/p ratio of H$_2$ is for our molecular cloud model; e.g. Flower et al. (2006) argued that the steady state value of the o/p ratio would not be attained before the protostellar collapse commences. Furuya et al. (in prep) recently calculated o/p ratio of H$_2$ in a molecular cloud formation. They showed that the o/p ratio of H$_2$ decreases to 10$^{-3}$, when the column density of a cloud reaches $\sim$3 mag. We thus calculate models with the initial o/p ratio of 0.01 and 0.001.

Figures 6a and 6b show the model calculation results for the three temperature cases (10 K, 30 K, and 100 K) with a density of 10$^6$ cm$^{-3}$ or 10$^4$ cm$^{-3}$, respectively.
As shown in Figures 6a and 6b, at a given time, the DNC/HNC abundance ratio is significantly higher in the high density model than in the low density model.
This is mainly due to the longer chemical timescale in the lower density model.
In addition, higher initial o/p ratio of H$_2$ molecules leads to lower D/H ratio, as seen in Figure 6. The o/p ratio may be higher in less dense regions (Pagani et al. 2011), which may contribute to relatively low deuterium fractionation in the low density regions.
Thus, the low DNC/HNC ratio measured by using single-dish observations is likely due to low density gas traced by the $J$=1--0 lines.

In Figure 6a, we can see that the DNC/HNC ratio observed toward the DNC peak is reproduced at $t$$\sim$10$^4$-10$^5$ yr at temperature of $T$$<$30 K, while the DNC/HNC ratio at the HN$^{13}$C peak is consistent with the model with $T$$\sim$100 K. The constant temperature model, however, would be unrealistic; the gas should have been cold initially, and then heated by the star formation.
Thus, we next consider a sudden temperature rise from 10 K to a given temperature at the time of 3$\times$10$^4$ yr or 10$^5$ yr, which simulates the birth of a protostar.
Figures 7b and 7c show the temporal variation of the D/H ratio after the temperature-rise with a gas density of 10$^6$ cm$^{-3}$ and the initial o/p ratio of 0.01; $t$=0 yr is defined as the time of the temperature-rise.

In Figure 7b, we can see that the DNC/HNC ratio does not decrease rapidly after the temperature rise, so that the DNC/HNC ratio toward the DNC peak could be reproduced with relatively high temperature (20-80 K) in the temperature-rise model.
After the temperature-rise, the main formation reaction of DNC is dissociative recombination of HCND$^+$, which is formed via DCO$^+$ + HCN. The DCO$^+$/HCO$^+$ ratio at warm temperature is enhanced by HCO$^+$ + D $\to$ DCO$^+$ + H in our model calculations.
Although deuterium enrichment of HNC via CH$_2$D$^+$ (e.g. Roueff et al. 2007) works in the early time of our constant temperature models, those reactions are negligible in the warm phase of the temperature-rise models; most of the carbon is already locked into CO at the time of temperature-rise and CH$_2$D$^+$ is not abundant.

We compare the observed DNC/HNC ratio at the HN$^{13}$C peak with the model results at t$\sim$10$^3$ yr, which is the dynamical timescale of the outflow (Paper I).
In Figure 7, we can see that the DNC/HNC ratio depends on the ratio at the temperature-rise.
If the temperature is raised at 1$\times$10$^5$ yr, the DNC/HNC ratio after the temperature rise is much higher than observed (Figure 7b).
The observed value toward the HN$^{13}$C peak is reproduced by the high temperature ($>$90 K) cases with the early temperature-rise ($<$3$\times$10$^4$ yr).
Thus, the low DNC/HNC ratio toward the HN$^{13}$C peak suggests that the temperature toward the HN$^{13}$C peak have risen in relatively early time, so that the D/H ratio in the gas and ice (see below) at $t$=0 is not too high.

In the model calculations, the low DNC/HNC ratio above 90 K originates from sublimation of grain surface molecules, whose DNC/HNC ratio at the cold phase (10 K) is lower than that in the gas phase, as seen in Figure 7a.
Figure 8a shows the DNC and HNC abundances in the ice mantle and in the gas phase for the constant temperature model (10 K), and Figure 8b for the temperature-rise models (from 10 K to 60 K or 100 K); the data plotted in Figures 8a and 8b are the same as those used in Figures 7a and 7c, respectively.
In Figure 8b, we can see that both HNC and DNC abundances in the gas phase are enhanced by more than one order of magnitude after the temperature rises to 100 K, and remain abundant for $>$10$^3$ yr.
If both HNC and DNC were sublimated from dust grains efficiently in the hot core, both HNC and DNC abundances should be peaked toward the hot core in the clump.
As mentioned before, this is not the case; the DNC peak is offset from the hot core.

One possibility to solve this contradiction is a low beam filling factor of the hot core (HN$^{13}$C peak); we proposed in paper I that the hot core consists of a few unresolved substructures.  Since DNC/HNC ratio in grain mantle is lower in that in the gas phase, DNC is not significantly enhanced toward the hot core unlike HN$^{13}$C.  This enhancement is further diluted by the beam filling factor in the observation.  Thus, the enhancement of DNC is hardly recognizable in comparison with the overwhelming DNC emission around the nearby DNC peak.  In order to investigate the low DNC abundance toward the hot core in more detail, higher angular resolution observations would be necessary. In addition, observations of DNC and HN$^{13}$C toward other hot cores would be also necessary for a full understanding.

\subsection{Implications from the DNC/HNC Ratio}

In this study, we have shown that the DNC/HNC abundance ratio is relatively high in the densest regions, while it is low in the diffuse envelope.
Taking these results into account, we give another possible interpretation of the single-dish survey made by Sakai et al. (2012).

Sakai et al. (2012) reported that the DNC/HNC ratio in high-mass sources is lower than that in low-mass sources.
One possible explanation is that the filling factor of the high density regions is different between observations of low-mass and high-mass objects.
Since the low-mass sources are typically much closer to earth than the high-mass sources, the filling factor of the high-density regions may be higher in the low-mass sources than in the high-mass sources.
In addition, it may also be possible that the DNC/HNC ratio of the high-mass sources is lower than that of the low-mass sources for a given density region (10$^4$ cm$^{-3}$) traced by the $J$=1--0 lines of HNC and DNC, for instance, due to higher temperature.
In order to investigate the difference between the low-mass and high-mass sources in more detail, multi-transition line observations with high angular resolution are therefore crucial.

Sakai et al. (2012) also reported that there is a diversity of the DNC/HNC ratio in the high-mass sources, where the DNC/HNC ratio is different even among the objects with similar evolutionary stages.
Note that there is no correlation between the DNC/HNC ratio and the distance.
A possible origin of this diversity is variation of the filling factor of high density regions; the objects with higher DNC/HNC ratio might contain more dense cores.
Another possibility is that the DNC/HNC ratio of the densest regions differs from source to source.
In this case, the ages of dense regions would be different from source to source.
In either case, the diversity of the DNC/HNC ratio could reflect the difference in the history of cluster formations.
Observations of the DNC/HNC ratio at a high spatial resolution have a high potential for elucidating the history of cluster formation.

\section{Summary}

We have determined the DNC, HN$^{13}$C and N$_2$H$^+$ distributions toward the IRDC clump G34.43+00.24 MM3 at high-angular resolution with ALMA.
The results are summarized below.

\begin{itemize} 

\item The N$_2$H$^+$ emission is anti-correlated with the CS emission.  Since N$_2$H$^+$ is destroyed by CO in warm gas, this anti-correlation indicates that the outflow, which is traced by CS, is interacting with the cold ambient gas.

\item The DNC and HN$^{13}$C emission is found to be distributed around the hot core. This suggests that both DNC and HN$^{13}$C emission comes from relatively warm regions ($>$ 20 K).

\item The DNC emission is stronger than the HN$^{13}$C emission toward many parts of this clump.  The DNC/HNC ratio averaged within a 15$^{\prime\prime}$$\times$15$^{\prime\prime}$ area around the phase center is higher than 0.06. This ratio is much higher than that in the previous single-dish observation ($\sim$0.003).  This discrepancy may result from the fact that the DNC/HNC ratio obtained by ALMA arises from relatively dense regions, which have only a small filling factor in the single dish observations.

\item We have compared the observations with simulations with the chemical model.  We confirm that the DNC/HNC ratio depends strongly on the density.  In addition, we suggest that the DNC/HNC ratio after the temperature-rise depends on the DNC/HNC ratio at the temperature-rise.

\item Taking the observational results into account, we give another interpretation of the single-dish results previously reported by Sakai et al. (2012). The low DNC/HNC ratio in the high-mass sources obtained by single-dish observations are due, at least in part, to the low filling factor of the high density regions. In addition, the diversity of the DNC/HNC ratio obtained by the single-dish observations among the high-mass sources may be due to the difference in the filling factor of high density regions or the evolutionary states of dense regions.

\end{itemize}

\acknowledgments

This paper makes use of the following ALMA data: ADS/JAO.ALMA\#2011.0.00656.S. 
ALMA is a partnership of ESO (representing its member states), NSF (USA) and NINS (Japan), together with NRC (Canada) and NSC and ASIAA (Taiwan), in cooperation with the Republic of Chile. 
The Joint ALMA Observatory is operated by ESO, AUI/NRAO, and NAOJ. We are grateful to the ALMA staffs. 
This study is supported by KAKENHI (21224002, 23740146, 24684011, 25400225 and 25108005).
JMJ acknowledges funding support from the US National Science Foundation via grant AST 1211844.
K.F. is supported by the Postdoctoral Fellowship for Research Abroad from the Japan Society
for the Promotion of Science (JSPS).

\clearpage



\begin{figure}
\epsscale{1.0}
\plotone{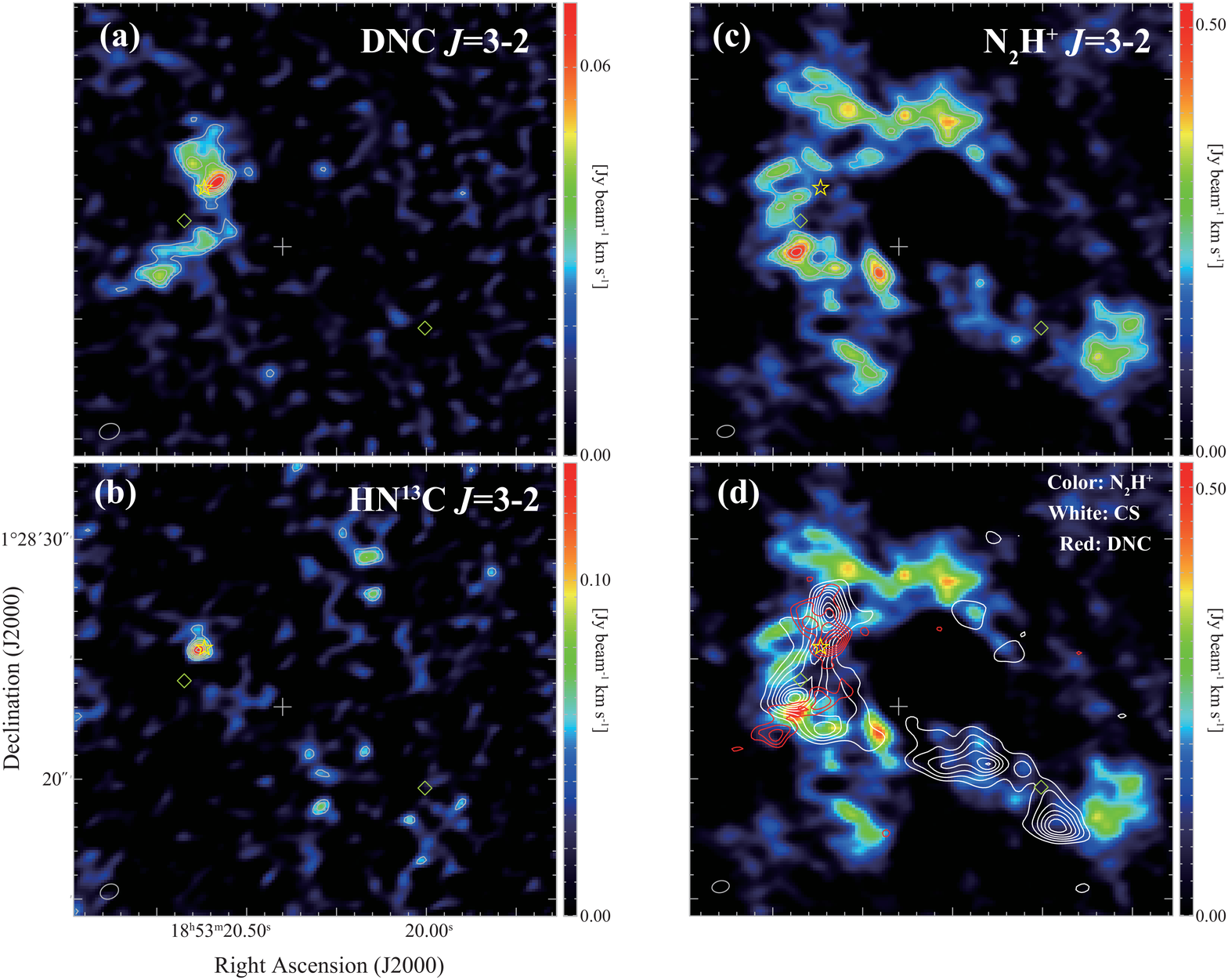}
\caption{Integrated intensity maps of DNC $J$=3--2 (a), HN$^{13}$C $J$=3--2 (b), and N$_2$H$^+$ $J$=3--2 (c) toward G34.43+00.24 MM3. Contour levels start from 3$\sigma$ and increase in steps of 1$\sigma$ [(a) $1\sigma=8$ mJy beam$^{-1}$ km s$^{-1}$, (b) $1\sigma=14$ mJy beam$^{-1}$ km s$^{-1}$. (c) $1\sigma=70$ mJy beam$^{-1}$ km s$^{-1}$]. Cross, star, and diamond marks indicate the positions of the phase center, the hot core, and the $Spizter$ sources (Shepherd et al. 2007), respectively. 
(d) The CS $J$=5--4 (Sakai et al. 2013) and HN$^{13}$C contours overlied on the N$_2$H$^+$ color image. 
\label{fig1}}
\end{figure}

\clearpage

\begin{figure}
\epsscale{1.0}
\plotone{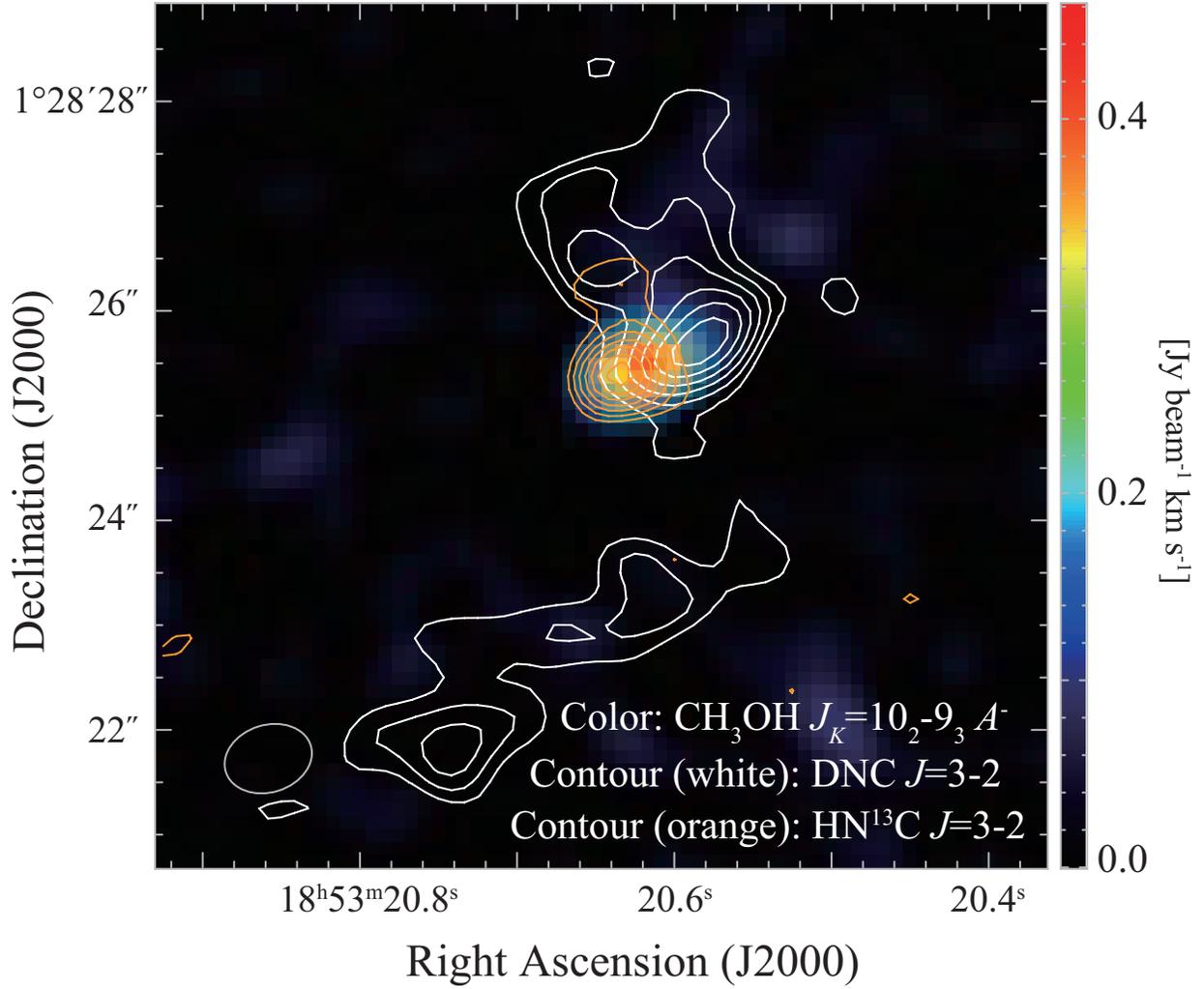}
\caption{Integrated intensity maps of DNC $J$=3--2 (white contours) and HN$^{13}$C $J$=3--2 (red contours) overlaid on that of CH$_3$OH $J_K$=10$_3$-9$_2$ $A^-$ (color; Sakai et al. 2013). Contour levels are the same as Figure 1. 
\label{fig2}}
\end{figure}

\clearpage

\begin{figure}
\epsscale{1.0}
\plotone{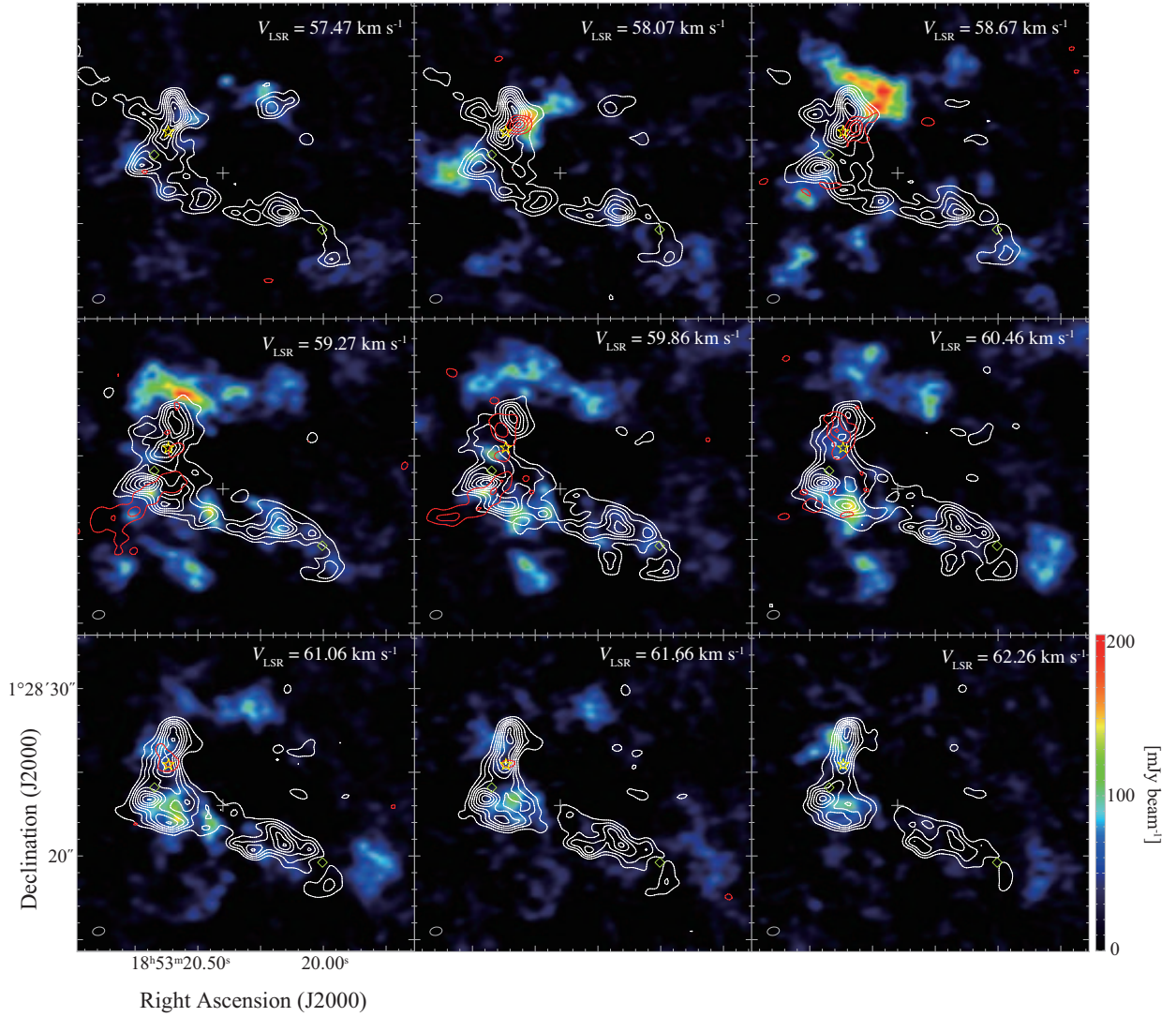}
\caption{Velocity-channel maps of the N$_2$H$^+$ (color), DNC (red) and CS (white) emission.  Contour levels start and increase in steps of 3$\sigma$ [DNC: $3\sigma=9$ mJy beam$^{-1}$ and CS: $3\sigma=42$ mJy beam$^{-1}$]. Cross, star, and diamond marks are the same as Figure 1.\label{fig3}}
\end{figure}


\clearpage

\begin{figure}
\epsscale{0.3}
\plotone{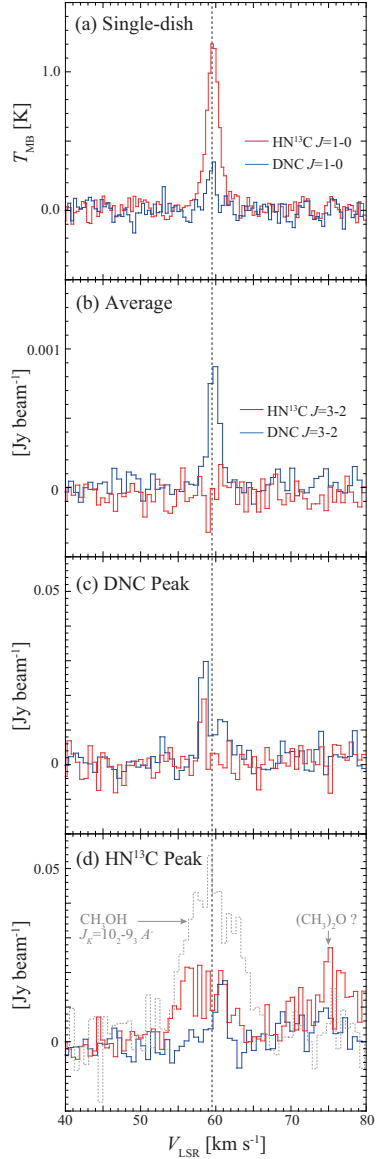}
\caption{(a) Spectra of DNC $J$=1--0 and HN$^{13}$C $J$=1--0 toward G34.43+00.24 MM3 obtained with the Nobeyama Radio Observatory 45 m telescope (Sakai et al. 2012).
(b) Spectra of DNC $J$=3--2 and HN$^{13}$C $J$=3--2 obtained with ALMA. The data are averaged within the 15$^{\prime\prime}$$\times$15$^{\prime\prime}$ region around the phase center.
(c) Spectra of DNC $J$=3--2 and HN$^{13}$C $J$=3--2 obtained with ALMA toward the DNC peak. (d) Same as (c), but toward the HN$^{13}$C peak. In (c), we plot the spectrum of CH$_3$OH $J_K$=10$_3$--9$_2$ $A^-$.
Vertical dotted lines indicates the peak velocity of the DNC $J$=1--0 emission (59.5 km s$^{-1}$).
\label{fig4}}
\end{figure}


\clearpage

\begin{figure}
\epsscale{1.0}
\plotone{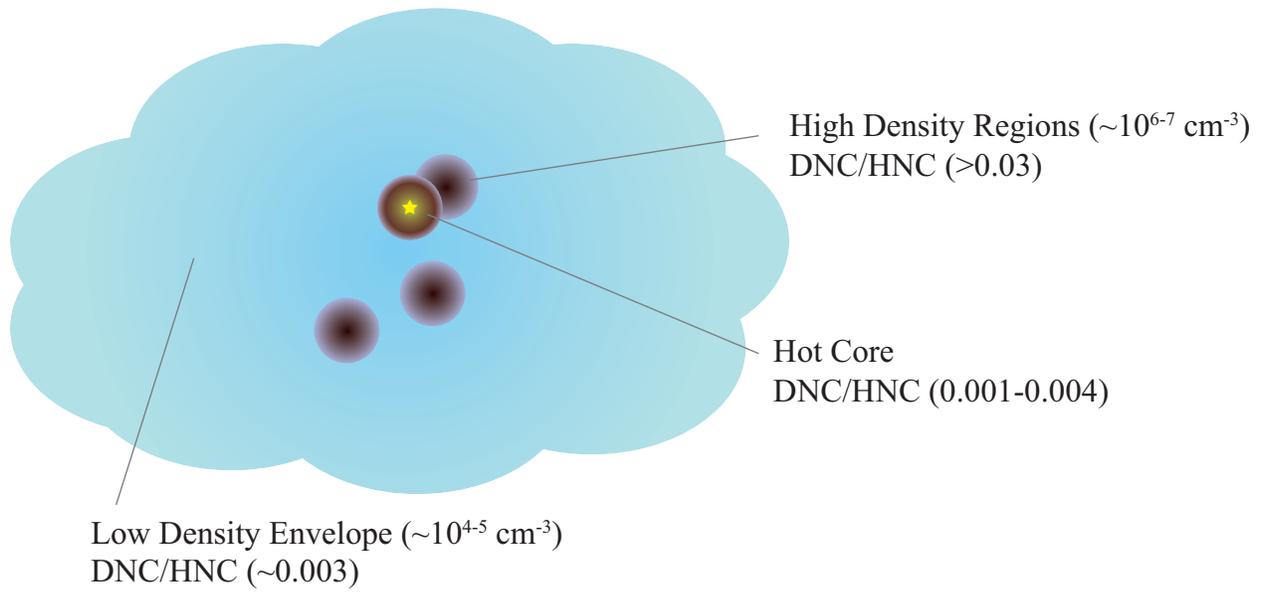}
\caption{A schematic illustration of the distribution of the DNC/HNC abundance ratio in a clump with embedded cores.
\label{fig5}}
\end{figure}


\clearpage

\begin{figure}
\epsscale{1.0}
\plotone{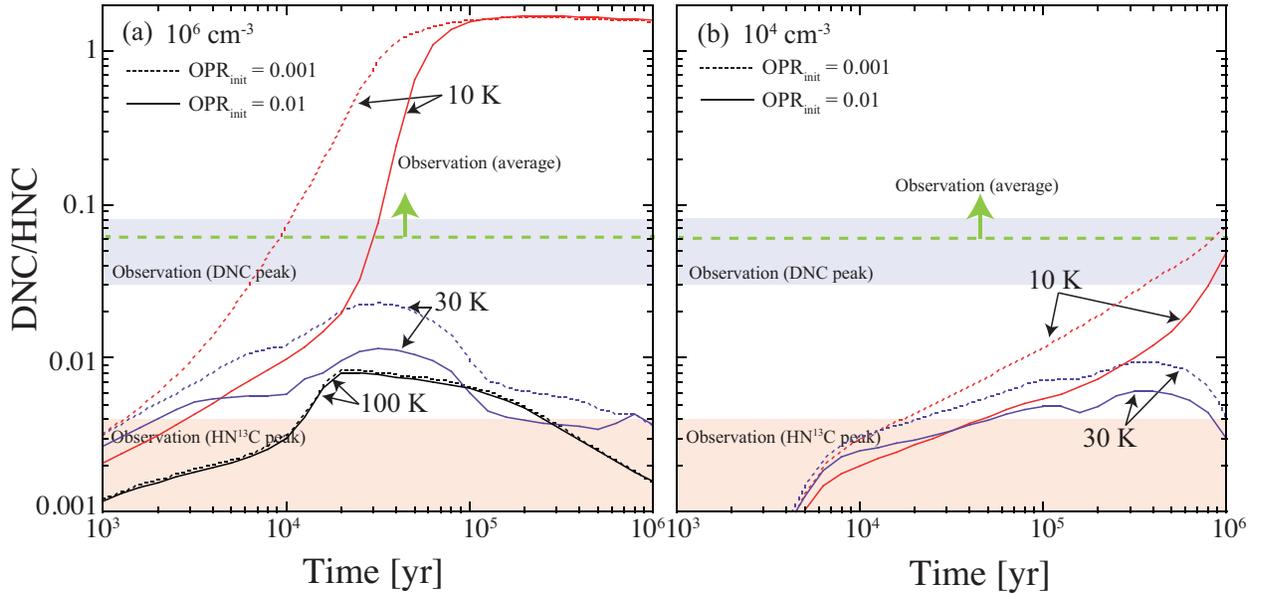}
\caption{(a) Chemical model calculation results of the deuterium fractionation ratio of DNC/HNC assuming the constant temperatures and the constant density of 10$^6$ cm$^{-3}$. The initial o/p ratio of H$_2$ molecules is assumed to be 0.01 (solid lines) or 0.001 (dashed lines).
The observed value for the DNC peak, the HN$^{13}$C peak, and the average spectra is plotted as purple region, pink region, and green dashed line, respectively. 
(b) Same as (a), but the density is 10$^{4}$ cm$^{-3}$.
\label{fig6}}
\end{figure}


\clearpage

\begin{figure}
\epsscale{1.0}
\plotone{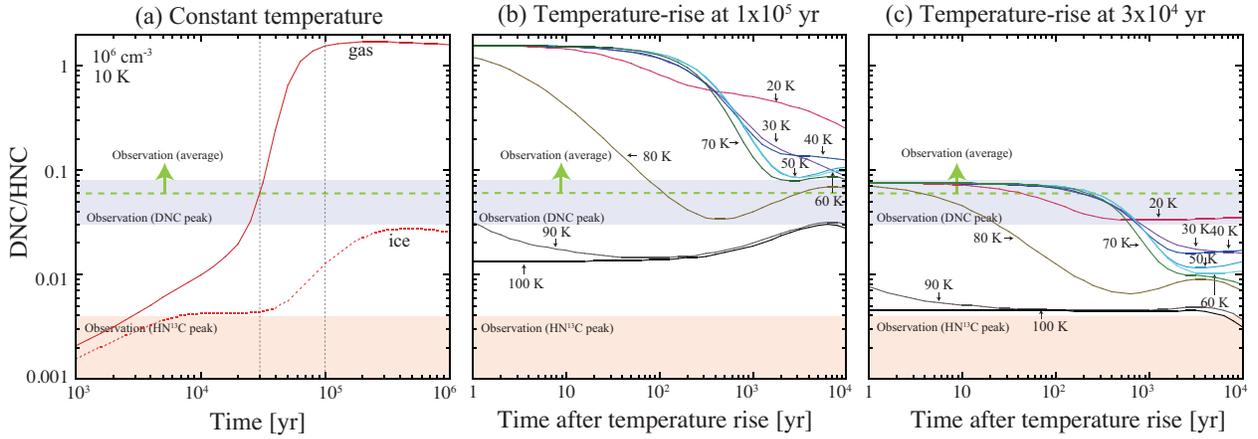}
\caption{(a) Chemical model calculation results of the DNC and HNC abundances in the gas phase and the ice mantle, assuming the constant temperature of 10 K, the constant density of 10$^6$ cm$^{-3}$ and the initial o/p ratio of H$_2$ molecules of 0.01.  Dashed vertical lines represent the time at which the temperature is raised for (b) and (c).
(b) Chemical model calculation results for the cases that the temperature is changed from 10 K to a given temperature (20-100 K) at the time of 10$^5$ yr in Figure 6a. 
(b) Same as (a), but the temperature rise at 3$\times$10$^{4}$ yr.
\label{fig7}}
\end{figure}


\clearpage

\begin{figure}
\epsscale{1.0}
\plotone{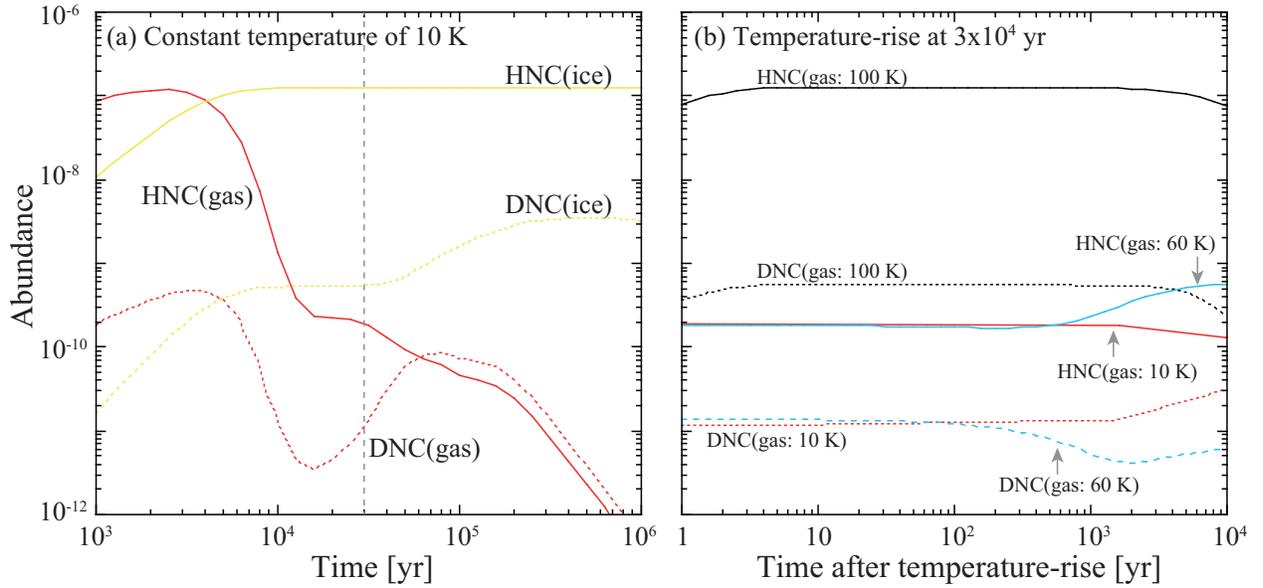}
\caption{Chemical model calculation results of the abundances of DNC and HNC relative to H$_2$.
(a) Constant temperature case of 10 K. The DNC and HNC abundances in the ice mantle are indicated by yellow lines, and those in the gas phase by red lines. Dashed vertical lines represent the time at which the temperature is raised for (b). (b) Temperature-rise cases, and the time is measured from the moment when the temperature is changed from 10 K to 60 K or 100 K. The DNC and HNC abundances in the gas phase are plotted.  In (b), we also plot the results of the constant temperature case of 10 K for comparison.
\label{fig8}}
\end{figure}


\clearpage







\clearpage

\begin{deluxetable}{ccrrrr}
\tabletypesize{\scriptsize}
\tablecaption{Observed lines\label{tbl-1}}
\tablewidth{0pt}
\tablehead{
\colhead{Species} & \colhead{Transition} & \colhead{Frequency} & \colhead{$E_u$/$k$} & \colhead{Beam size} & \colhead{P.A.} \\
 & & \footnotesize[GHz]  & \footnotesize[K] &  \footnotesize[arcsec] & \footnotesize[degree]\\
}
\startdata
DNC & $J$=3--2 & 228.91048 & 22.0 & 0.85$\times$0.64 & -72 \\
HN$^{13}$C & $J$=3--2 & 261.26351 & 25.1 & 0.79$\times$0.60 & -65 \\
N$_2$H$^+$ & $J$=3--2 & 279.51173 & 26.8 & 0.73$\times$0.49 & 101 \\
\enddata
\end{deluxetable}

\clearpage
\begin{deluxetable}{lrrrrrr}
\rotate
\tablecolumns{7} 
\tablewidth{0pc} 
\tabletypesize{\small} 
\tablecaption{DNC/HNC Abundance Ratio.} 
\tablehead{ 
\colhead{Position} & \colhead{R. A.}   & \colhead{Dec.}    & \colhead{$I_{\rm DNC}$}   & \colhead{$I_{\rm HN^{13}C}$}  & \colhead{$N$(DNC)} & \colhead{$N$(DNC)/$N$(HNC)\tablenotemark{a}} \\
 & \footnotesize(J2000.0) & \footnotesize(J2000.0)  & \footnotesize[mJy beam$^{-1}$ km s$^{-1}$] &  \footnotesize[mJy beam$^{-1}$ km s$^{-1}$] & \footnotesize{10$^{12}$ [cm$^{-2}$]}  & \\
 }
\startdata 
DNC Peak & 18 53 20.58 & 1 28 25.6 & 79$\pm$3 & 24$\pm$5 & 2.8-5.3\tablenotemark{b} & 0.03-0.08\tablenotemark{b} \\
HN$^{13}$C Peak & 18 53 20.63 & 1 28 25.4 & 17$\pm$4 & 130$\pm$4 & 0.48-2.0\tablenotemark{c} & 0.001--0.004\tablenotemark{c} \\
Averaged Spectra & --- & --- & 1.65$\pm$0.08 & $<$0.3 & 0.058--0.11\tablenotemark{b} & $>$0.06\tablenotemark{b} \\
\enddata 
\tablenotetext{a}{The HN$^{12}$C/HN$^{13}$C ratio is assumed to be 60-80.}
\tablenotetext{b}{The excitation temperature is assumed to be 20--80 K.}
\tablenotetext{c}{The excitation temperature is assumed to be 20--130 K.}
\label{tab:t2}
\end{deluxetable} 

\clearpage


\clearpage




\end{document}